\documentclass[
 reprint,]{revtex4-2}

\usepackage{graphicx}
\usepackage{dcolumn}
\usepackage{bm}
\usepackage{xcolor}
\usepackage{cancel}
\newcommand{\ket}[1]{| {#1} \rangle} 
\newcommand{\bra}[1]{\langle {#1} |} 

\usepackage[utf8]{inputenc}
\usepackage[T1]{fontenc}
\usepackage[english,serbian]{babel}
\usepackage[OT1]{fontenc}
\usepackage{mathrsfs}
\usepackage{graphicx}
\usepackage{amssymb}
\usepackage{amsbsy}
\usepackage{latexsym}
\usepackage{mathtools}
\usepackage{titlesec}
\usepackage{amsmath}
\usepackage{color}
\usepackage[utf8]{inputenc}
\usepackage{lipsum}
\usepackage{fontenc}

\usepackage{chngcntr}

\def\nn{\nonumber}

\def\de{\textrm{d}}

\def\dj{d\kern-0.4em\char"16\kern-0.1em}
\def \Dj {\mbox{\raise0.3ex\hbox{-}\kern-0.4em D}}

\begin{document}

\preprint{APS/123-QED}

\title{Page Curve for Eternal Schwarzschild Black Hole in Dimensionally-Reduced Model of Dilaton Gravity}

\author{Stefan \Dj or\dj evi\'{c}, Aleksandra Go\v{c}anin, Dragoljub Go\v{c}anin and Voja Radovanovi\'{c}}
\affiliation{Faculty of Physics, University of Belgrade, Studentski Trg 12-16, 11000 Belgrade, Serbia}

\selectlanguage{english}

\date{\today}

\begin{abstract}

As a contribution to the subject of the information loss paradox in (1+1)-dimensional gravitational systems, we study a model of (1+1)-dimensional dilaton gravity derived from the four-dimensional Einstein-Hilbert action by dimensional reduction. The reduced action involves the cosmological constant and admits black hole solutions.
After including the back-reaction of quantum fields to 1-loop order, we solve the semi-classical field equations perturbatively and compute the quantum correction to the Hawking temperature. We consider the quantum extremal surface approach and invoke the ``island rule'' to compute the fine-grained entropy of the Hawking radiation for an eternal Schwarzschild black hole and demonstrate that it follows the unitary Page curve. 

\end{abstract}

\maketitle

\selectlanguage{english}

\section{Introduction}

It is widely believed that the resolution of the information loss paradox \cite{Hawking1} could be the key to gaining a deeper insight into the quantum nature of gravity and, in particular, the microscopic origin of the black hole entropy \cite{Bekenstein1, Bekenstein2, Hawking2}. The standard calculation done by Hawking \cite{Hawking3} predicts that, due to the thermal character of the Hawking radiation, the process of black hole formation and evaporation breaks the principle of unitary time evolution and leads to a monotonic increase of the fine-grained entanglement entropy of the radiation even beyond the Bekenstein-Hawking (BH) entropy limit, $S_{\text{BH}}=A(\text{horizon})/4G^{(4)}_{\text{N}}$ (we set $c=\hbar=k_{\text{B}}=1$, and $G^{(4)}_{\text{N}}$ stands for the Newton's constant in four space-time dimensions), which, on the other hand, monotonically decreases due to evaporation. 

Various models of (1+1)-dimensional dilaton gravity, such as Jackiw-Teitelboim (JT) \cite{JT1, JT2} and Callan–Giddings–Harvey–Strominger (CGHS) \cite{CGHS}, have proved to be very useful for analytical investigation of the process of black hole formation and evaporation. This is due to the fact that, after integrating out fluctuations of the matter fields and including $1$-loop quantum corrections, field equations can be made exactly solvable by introducing suitable correction terms, as in Russo-Susskind-Thorlacius (RST), Bose-Parker-Peleg (BPP) and CGSH model \cite{RST1, RST2, BPP, QCGHS}. The results suggest that information does indeed get lost in the process of black hole evaporation, signalling the breakdown of unitarity. A more comprehensive account of 
dilaton gravity models can be found in \cite{Fabri, DWV}.         
       
However, the requirement of information conservation implies that the entanglement entropy of the quantum fields outside the black hole should not exceed the course-grained limit set by $S_{\text{BH}}$ and 
must follow the so-called Page curve instead \cite{Page1, Page2}. 
How this kind of behaviour might arise is a topic that has received a lot of attention recently. 

One proposal that stands out is based on the idea that the fine-grained entropy of the Hawking radiation can receive an extra contribution from the so-called ``island'' \cite{Penington, AEMM, AMMZ, AMM}. The state of Hawking radiation, corresponding to a spatial region $R$ outside the black hole, is standardly described by a density matrix obtained by taking a partial trace over the degrees of freedom in the complementary region $\Bar{R}$; an island $I\subset\bar{R}$ is supposed to be a part of $\Bar{R}$ that should be excluded when taking the partial trace. According to the minimal
quantum extremal surface (QES) prescription \cite{RT, HRT, QES}, the fine-grained (FG) entropy of the Hawking radiation corresponding to $R$ is given by the ``island formula'',
\begin{equation}\label{Island_formula}
S_{FG}(R)=\underset{I}{\min}\left\{\underset{I}{\text{ext}}\left[\frac{\text{A}(\partial I)}{4G_{N}}+S_{\text{matter}}(R\cup I)\right]\right\},
\end{equation}
where $S_{\text{matter}}(R\cup I)$ stands for the semi-classical entanglement entropy of the quantum fields with support on $R\cup I$, and $A(\partial I)$ is the area of the $I$'s boundary surface (it need not be the event horizon). The island is a surface that extremizes the generalized entropy functional in the square brackets of (\ref{Island_formula}). If there are several such quantum extremal surfaces, the prescription dictates that we should pick the one that minimizes the generalized entropy.

Apart from its holography origins, the ``island rule'' gains additional support from considerations regarding the gravitational path integral. Namely, the replica trick \cite{CW, HLW, CC} applied to gravitational
systems leads to a new kind of saddle-points, the so-called ``replica wormholes'', for which bulk wormholes are connecting different copies of space-time. These new saddle-point configurations give rise to islands \cite{PSSY, AHMST}, and the unitary Page curve comes as a non-perturbative effect from replica wormholes.

In the semi-classical limit, the partition function of the geometry with replicas is dominated by the one giving the minimum entropy. In this way, the replica trick for gravitational theories gives rise to the same formula (\ref{Island_formula}) as the holographic QES prescription.  
Since the replica wormhole is merely a consequence of the replica trick in models with gravitation, the island rule is expected to be applicable to any kind of black hole. So far, the island rule has been studied mainly in (1+1)-dimensional models, which offer a tractable treatment of the entanglement entropy of the Hawking radiation \cite{Penington, AEMM, AMMZ, AMM, PSSY, AHMST, CFHMR, Chen, AEPU, LV, MM, BKPSU, Bhattacharya, Verlinde, CQZ, GSST, APS, GHT, AI}, but also in higher-dimensional models \cite{AMS, HIM, WLW, Q}. 

The paper is organized as follows. In the following section, we study, in some detail, a model of (1+1)-dimensional dilaton gravity derived from Einstein-Hilbert action by dimensional reduction. 
Section III 
analyzes the contribution of $1$-loop quantum corrections to the energy-momentum tensor for matter fields. Focusing on the eternal Schwarzschild black hole solution, we compute quantum corrections to the metric, the position of the horizon and surface gravity in Section IV. Finally, in Section V, we derive the Page curve using the island rule. A conclusion and some proposals for future work are given in Section VI.    

\section{DREH model}

The DREH (Dimensionally-Reduced Einstein-Hilbert) model is a (1+1)-dimensional model of dilaton gravity obtained from the usual four-dimensional Einstein-Hilbert (EH) action by using a spherically symmetric ansatz and integrating out the angles. It is similar in some ways to the CGHS model of dilaton gravity \cite{CGHS}. In particular, it admits black hole solutions. 

Dimensional reduction is a well-known procedure, and we only give a brief review. More technical details can be found in \cite{MVM}. Start with the standard EH action in four space-time dimensions,  
\begin{equation}
S_{\text{EH}}=\frac{1}{16\pi G_{\text{N}}^{(4)}}\int \de^{4}x\sqrt{-g^{(4)}}R^{(4)},
\end{equation}
where $G_{\text{N}}^{(4)}$ is the four-dimensional Newton's constant, $g^{(4)}_{AB}$ $(A,B=0,1,2,3)$ is the four-dimensional metric, and $R^{(4)}$ is the four-dimensional curvature scalar.

Consider the following spherically symmetric ansatz for the metric, 
\begin{equation}\label{ansatz}
ds^{2}=g_{\mu\nu}d x^{\mu}d x^{\nu}+\lambda^{-2}e^{-2\phi}\left[d\theta^{2}
+\sin^{2}\theta d\varphi^{2}\right],     
\end{equation}
where $\mu,\nu=0,1$, and $g_{\mu\nu}$ depends only on $x^{0}$ and $x^{1}$. The dilaton field $\phi$ is related to the radial coordinate $r=\lambda^{-1}e^{-\phi}$, and $\lambda^{2}$ is a constant parameter that will play the role of the cosmological constant in the reduced theory. 

Using the ansatz (\ref{ansatz}), we can straightforwardly derive the relation between the curvature scalar $R$ of the reduced (1+1)-dimensional theory, and the curvature scalar $R^{(4)}$ of the four-dimensional theory,
\begin{align}
R^{(4)}=R+2(\nabla\phi)^{2}+2\lambda^{2}e^{2\phi}-2e^{2\phi}\Box e^{-2\phi},
\end{align}
Also, we have
\begin{align}
\de^{4}x\sqrt{-g^{(4)}}=\de^{2}x\de\theta\de\varphi\sqrt{-g}\frac{e^{-2\phi}}{\lambda^{2}}\sin^{2}{\theta}.
\end{align}
The reduced dilaton gravity action (up to a surface term), which will be denoted by $S_{\phi}$, comes down to
\begin{equation}\label{Dilaton_action}
S_{\phi}=\frac{1}{4G_{\text{N}}}\int \de^{2}x\sqrt{-g}\Big[e^{-2\phi}\left(R+2(\nabla\phi)^{2}\right)+2\lambda^{2}\Big],    
\end{equation}
where we introduced $G_{\text{N}}\equiv\lambda^{2}G_{\text{N}}^{(4)}$ as the Newton's constant of the reduced theory. 

Later we will take quantum corrections into account, and for that purpose we introduce the conformal matter term, $S_{m}$, for a massless scalar field $f$ minimally coupled to gravity. Therefore, the final form of the classical DREH action is given by
\begin{align}
&S_{\text{DREH}}=S_{\phi}+S_{m}\nn\\
&=\frac{1}{4G_{\text{N}}}\int \de^{2}x\sqrt{-g}\Big[e^{-2\phi}\left(R+2(\nabla\phi)^{2}\right)+2\lambda^{2}\Big]\nn\\
&-\frac{1}{2}\int \de^{2}x\sqrt{-g}\left(\nabla f\right)^{2}.\label{dejstvo}
\end{align}
Classical field equations are obtained by varying $S_{DREH}$ with respect to $g_{\mu\nu}$, $\phi$ and $f$, yielding, respectively,
\begin{widetext}
\begin{align} 
\left[2\nabla_{\mu}\nabla_{\nu}\phi-2\nabla_{\mu}\phi\nabla_{\nu}\phi +g_{\mu\nu}\left(3(\nabla\phi)^{2}-2\Box\phi-\lambda^{2}e^{2\phi}\right)\right] e^{-2\phi}&=2G_{\text{N}}T^{(f)}_{\mu\nu,\text{class}}, \label{g-equation}\\
            (\nabla\phi)^{2}-\Box\phi&=\frac{R}{2},\label{phi-equation}\\
            \Box f&=0.
\end{align}
\end{widetext}
with the classical energy-momentum tensor for the matter field $f$,
\begin{equation}
T^{(f)}_{\mu\nu,\text{class}}=\frac{-2}{\sqrt{-g}}\frac{\delta S_{m}}{\delta g^{\mu\nu}}=\nabla_{\mu}f\nabla_{\nu}f-\frac{1}{2}g_{\mu\nu}(\nabla f)^{2}.
\end{equation}
 
Now we will show that the vacuum solution of the classical field equations is in fact a Schwarzschild black hole. Consider the following static ansatz for the metric in $(t,r)$ coordinates,
\begin{equation} \label{classical_metric}
ds^{2}=-h_{0}(r)dt^{2}+\frac{dr^{2}}{h_{0}(r)},
\end{equation}
and $r=\lambda^{-1}e^{-\phi}$. The curvature scalar for this metric is $R=-\partial^{2}_{r}h_{0}$, and the dilaton equation (\ref{phi-equation}) becomes 
\begin{align}
\partial^{2}_{r}h_{0}+\frac{2}{r}\partial_{r}h_{0}=0,
\end{align}
which gives us the form of the unknown function $h_{0}(r)$, up to two undetermined constants,
\begin{equation}
h_{0}(r)=A-\frac{C}{r}.    
\end{equation}
Additionally, from (\ref{g-equation}) in the $rr$-case we get \begin{align}
h_{0}(r)=1-r\partial_{r}h_{0}(r),
\end{align}
implying that $A=1$ and leaving $C$ undetermined; the $tt$-equation does not impose any additional constraint. Thus we obtained a Schwarzschild black hole solution in two dimensions. In four dimensions, the constant $C$ is related to the mass of the black hole, $C=2MG_{\text{N}}^{(4)}$; therefore, in two dimensions we have $C=\frac{2MG_{\text{N}}}{\lambda^{2}}$. The full metric is given by
\begin{equation}
ds^{2}=-\left(1-\frac{2MG_{\text{N}}}{\lambda^{2}r}\right)dt^{2}+\frac{dr^{2}}{1-\frac{2MG_{\text{N}}}{\lambda^{2}r}}.
\end{equation}
From now on, we will use the notation $r_{0}=\frac{2MG_{\text{N}}}{\lambda^{2}}$ for the classical horizon radius, so that $h_{0}(r)=1-\frac{r_{0}}{r}$. 


\section{Quantum corrections in the DREH model}

Having established the classical DREH model, we 
consider the quantization of matter fields (a single massless scalar field in this case) on the classical background geometry of the Schwarzschild black hole. Quantum corrections come in the form of Polyakov-Liouville (PL) action \cite{Polyakov}, 
\begin{align}\label{SPL}
S_{\text{PL}}=-\frac{\hbar}{96\pi}\int\de^{2}x\int&\de^{2}x'\sqrt{-g(x)}\sqrt{-g(x')}\nonumber \\
&\times R(x)G(x-x')R(x'),
\end{align}
where $G(x-x')$ stands for the Green function for the massless Klein-Gordon equation in curved (1+1)-dimensional space-time. This action represents a $1$-loop effective action obtained by integrating out fluctuations of the massless scalar field, 
\begin{equation}
e^{\frac{i}{\hbar}S_{\text{PL}}}=\int\mathcal{D}\chi e^{\frac{i}{\hbar}\int\de^{2}x\sqrt{-g}[-\frac{1}{2}(\nabla \chi)^{2}]},   
\end{equation}
and it can be converted into a local form by introducing an auxiliary field $\psi$,
\begin{equation}
S_{\text{PL}}=-\frac{\hbar}{96\pi}\int\de^{2}x\sqrt{-g}\left[2R\psi+\left(\nabla\psi\right)^{2}\right],    
\end{equation}
which is on-shell equivalent to $(\ref{SPL})$, the field equation of the auxiliary field being 
\begin{equation}
\Box\psi=R.    
\end{equation}

The full action for the $1$-loop quantum DREH model is given by
\begin{equation}
S=S_{DREH}+S_{\text{PL}}.    
\end{equation}
 
Variation of $S_{\text{PL}}$ in terms of $g_{\mu\nu}$ gives us a quantum correction to the energy-momentum tensor for the scalar field $f$,
\begin{align} \label{T_1loop}
\langle\Delta T_{\mu\nu}^{(f)}\rangle&=\frac{-2}{\sqrt{-g}}\frac{\delta S_{\text{PL}}}{\delta g^{\mu\nu}}
=\frac{\hbar}{48\pi} \Bigg[-2\nabla_{\mu}\nabla_{\nu}\psi\nonumber\\
&+\nabla_{\mu}\psi\nabla_{\nu}\psi
+g_{\mu\nu}\left(2\Box\psi-\frac{1}{2}(\nabla\psi)^{2}\right)\Bigg].
\end{align}
To define $\langle\Delta T_{\mu\nu}^{(f)}\rangle=\langle\Psi\vert\Delta T_{\mu\nu}^{(f)}\vert\Psi\rangle$ we also need to specify the quantum state $\vert\Psi\rangle$ that we are considering.  
The full energy-momentum tensor consists of the classical part and the $1$-loop quantum correction coming from the PL effective action,
\begin{equation}
T_{\mu\nu}^{(f)}=T_{\mu\nu,class}^{(f)}+\langle\Delta T_{\mu\nu}^{(f)}\rangle.    
\end{equation}

The metric equation (\ref{g-equation}) changes only due to this quantum correction of the energy-momentum tensor,
\begin{widetext}
\begin{align} 
\Big[2\nabla_{\mu}\nabla_{\nu}\phi-2\nabla_{\mu}\phi\nabla_{\nu}\phi +g_{\mu\nu}\Big(3(\nabla\phi)^{2}-2\Box\phi&-\lambda^{2}e^{2\phi}\Big)\Big] e^{-2\phi}=2G_{\text{N}}\Bigg\{\nabla_{\mu}f\nabla_{\nu}f-\frac{1}{2}g_{\mu\nu}(\nabla f)^{2}\nn\\
&+\frac{\hbar}{48\pi} \Bigg[-2\nabla_{\mu}\nabla_{\nu}\psi
+\nabla_{\mu}\psi\nabla_{\nu}\psi
+g_{\mu\nu}\left(2\Box\psi-\frac{1}{2}(\nabla\psi)^{2}\right)\Bigg]\Bigg\}.
\end{align}
\end{widetext}

On the other hand, the dilaton equation (\ref{phi-equation}) remains the same because we treat $\phi$ as a purely classical field.   

The DREH model is derived directly from the physically relevant EH action. However, an important aspect of this model is that it can not be made exactly solvable by adding suitable correction terms in the action, and one has to use the perturbation technique as in \cite{MVM, MV, Frolov}. Quantity $\epsilon=\frac{\hbar}{48\pi}$ is a natural perturbation parameter.  

The classical metric is given by $(\ref{classical_metric})$ with $h_{0}(r)=1-r_{0}/r$. We are interested in finding an eternal black hole solution at one-loop order. For that we introduce two functions $\varphi(r)$ i $m(r)$, while keeping the definition $r=\lambda^{-1}e^{-\phi}$, and assume the following static ansatz for the metric,
\begin{equation}
ds^{2}=-h(r)e^{2\epsilon\varphi(r)}dt^{2}+\frac{dr^{2}}{h(r)},
\end{equation}
with 
\begin{equation}
h(r)=h_{0}(r)+\frac{\epsilon m(r)}{r}=1-\frac{r_{0}}{r}+\frac{\epsilon m(r)}{r}.
\end{equation}
The curvature scalar is
\begin{equation}\label{R-quantum}
R=-\partial^{2}_{r}h-2\epsilon h\partial^{2}_{r}\varphi-3\epsilon\partial_{r}h\partial_{r}\varphi-2\epsilon^{2}h(\partial_{r}\varphi)^{2}.    
\end{equation}
From the modified field equations for the metric (which now include the back-reaction to $1$-loop order) we get 
\begin{align}\label{diff_equations_m}
\frac{dm}{dr}&=-\frac{\tilde{T}^{(f)}_{tt}}{h_{0}},\\
2rh_{0}\frac{d\varphi}{dr}&=h_{0}\tilde{T}^{(f)}_{rr}+\frac{\tilde{T}^{(f)}_{tt}}{h_{0}},\label{diff_equations_varphi}
\end{align}
where we introduced $\tilde{T}^{(f)}_{\mu\nu}=\frac{2G_{\text{N}}}{\lambda^{2}\epsilon}T^{(f)}_{\mu\nu}$. The dilaton equation reduces to an identity, and it has no bearing for further analyses. 

Since components of the energy-momentum tensor $T^{(f)}_{\mu\nu}$ are already of order $\epsilon$, we can use classical metric components in 
this perturbative order. First, we need to solve the $\Box\psi=R$ equation perturbatively for the auxiliary field. 
From the result (\ref{R-quantum}) we get
\begin{equation}
\partial_{r}\left(h\partial_{r}\psi+\partial_{r}h\right)=0,
\end{equation}
and, therefore
\begin{equation}
\partial_{r}\psi=\frac{C-\partial_{r}h}{h},\label{1}   
\end{equation}
where $C$ is an undetermined constant that is related to the quantum state of the radiation.

We can now use (\ref{T_1loop}) to compute the quantum corrections to the energy-momentum tensor up to first order in $\epsilon$. They are given by  
\begin{align}\label{T_Sch_rr}
\langle\Delta T^{(f)}_{rr}\rangle&=\epsilon\frac{C^{2}-(\partial_{r}h_{0})^{2}}{2h_{0}^{2}},\\
\langle\Delta T^{(f)}_{tt}\rangle&=\epsilon\left[\frac{C^{2}}{2}-\frac{1}{2}(\partial_{r}h_{0})^{2}+2h_{0}\partial^{2}_{r}h_{0}\right],\label{T_Sch_tt}
\end{align}
up to an undetermined constant $C$.

\section{Eternal black hole scenario}

Now we are ready to consider the eternal Schwarzschild black hole solution, where an incoming
energy flux balances energy loss due to Hawking evaporation. For a distant observer (i.e. in asymptotically flat coordinates), the black hole appears to be in thermal equilibrium with its environment, which corresponds to the Hartle-Hawking (HH) state of the radiation. On the other hand, in Kruskal coordinates $x^{\pm}$, the $\vert HH\rangle$ state represents the vacuum. This condition will determine the value of the integration constant $C$ mentioned earlier.

First we compute the $\epsilon$-correction to the horizon's position, $r_{H}=r_{0}+\epsilon r^{(1)}$, as determined by the condition
\begin{equation}
g^{rr}(r_{H})=0, \hspace{0.2cm}\text{i.e.}\hspace{0.2cm} h(r_{H})=1-\frac{r_{0}}{r_{H}}+\frac{\epsilon m(r_{H})}{r_{H}}=0, 
\end{equation}
yielding
\begin{equation}
r^{(1)}=-m(r_{0}).     
\end{equation}
Therefore, the new position of the horizon is 
\begin{equation}
r_{H}=r_{0}-\epsilon m(r_{0}).
\end{equation}

To make a transition to Kruskal coordinates, we have to find the quantum correction to the surface gravity. The redshift factor is 
\begin{equation}
V=\sqrt{-g_{\mu\nu}\xi^{\mu}\xi^{\nu}}=\sqrt{-g_{tt}}=\sqrt{he^{2\varepsilon\varphi}},
\end{equation}
and the surface gravity (note that $h(r_{H})=0$) is
\begin{align}
    \kappa&=\sqrt{(\nabla V)^{2}}=\sqrt{g^{rr}}\partial_{r}V=\frac{1}{2}\partial_{r}h e^{\epsilon\varphi}\\
    &=\frac{1}{2r_{0}}\left[1+\epsilon\left(\varphi(r_{0})+\partial_{r}m(r_{0})+\frac{m(r_{0})}{r_{0}}\right) +\mathcal{O}(\epsilon^{2})\right].\nonumber
\end{align}
The metric can be represented in a conformal form,
\begin{align}
ds^{2}&=-h(r)e^{2\epsilon\varphi(r)}dt^{2}+\frac{dr^{2}}{h(r)}\nonumber \\
&=h(r)e^{2\epsilon\varphi(r)}\left(-dt^{2}+\frac{dr^{2}}{h^{2}(r)e^{2\epsilon\varphi(r)}}\right)\nonumber \\
&=h(r)e^{2\epsilon\varphi(r)}\left(-dt^{2}+dr^{2}_{*}\right),
\end{align}
where we introduce the tortoise coordinate
\begin{equation}
r_{*}=\int\frac{e^{-\epsilon\varphi(r)}}{h(r)}dr.
\end{equation}
Since $h(r_{H})=0$, this integral diverges at $r=r_{H}$. Note also that $\kappa=\frac{1}{2}\left(\frac{dh}{dr}e^{\epsilon\varphi}\right)\bigg{|}_{H}$, which means that $\frac{dh(r_{H})}{dr}\neq0$. By expanding the integral near the horizon $r_{H}$ we get
\begin{align}
    r_{*}&=\frac{1}{2\kappa}\int\left[\frac{1}{r-r_{H}}-\left(\frac{1}{2}\frac{h''}{h'}+\epsilon\frac{d\varphi}{dr}\right)\bigg{|}_{H}+o(r-r_{H})\right]dr\nonumber\\
    &=\frac{1}{2\kappa}\left[\frac{r}{r_{H}}+\ln{\left(\frac{r}{r_{H}}-1\right)}+\epsilon\alpha(r)\right],\label{r_zvezda}
\end{align}
where $\epsilon\alpha(r)$ represents the part of the integral that does not diverge at the horizon, and $\alpha=\alpha^{(0)}+\epsilon\alpha^{(1)}+\mathcal{O}(\epsilon^{2})$. Therefore, to first order in $\epsilon$ we can write
\begin{equation}
r_{*}=\int\frac{1-\epsilon\left(\varphi(r)+\frac{m(r)}{r-r_{0}}\right)}{1-\frac{r_{0}}{r}}\de r.   
\end{equation}

In asymptotically flat coordinates $\sigma^{\pm}=t\pm r_{*}$, the metric becomes
\begin{equation}
ds^{2}=-h(r)e^{2\epsilon\varphi(r)}d\sigma^{+}d\sigma^{-}.
\end{equation}
Finally, in Kruskal coordinates, $\kappa x^{\pm}=\pm e^{\pm\kappa\sigma^{\pm}}$, it can be represented as
\begin{align}
ds^{2}&=-he^{2\epsilon\varphi}\frac{d\sigma^{+}}{dx^{+}}\frac{d\sigma^{-}}{dx^{-}}dx^{+}dx^{-}\nonumber\\
&=-he^{2\epsilon\varphi}\frac{dx^{+}dx^{-}}{-\kappa^{2}x^{+}x^{-}}\nonumber \\
&=-he^{2\epsilon\varphi-2\kappa r_{*}}dx^{+}dx^{-}=-e^{2\rho}dx^{+}dx^{-},
\end{align}
with the conformal factor
\begin{equation}
\rho(x)=\frac{1}{2}\ln{h}+\epsilon\varphi-\kappa r_{*}.\label{2}
\end{equation}
The equation $\Box\psi=R$ for the auxiliary field in Kruskal coordinates is simply $\partial_{+}\partial_{-}(\psi+2\rho)=0$, which is solved by $\psi=-2\rho+F_{+}(x^{+})+F_{-}(x^{-})$, where we introduced two arbitrary functions $F_{\pm}(x^{\pm})$. 

The quantum corrections to the energy-momentum tensor in $x^{\pm}$ coordinates are given by
\begin{align} \label{T_Kruskal}
    \langle\Delta T^{(f)}_{\pm\pm}\rangle&=4\epsilon\left[\partial^{2}_{\pm}\rho-\left(\partial_{\pm}\rho\right)^{2}-t_{\pm}(x^{\pm})\right],\\
    \langle\Delta T^{(f)}_{+-}\rangle&=-4\epsilon\partial_{+}\partial_{-}\rho,
\end{align} 
where $t_{\pm}(x^{\pm})=\frac{1}{2}\partial_{\pm}^{2}F_{\pm}-\frac{1}{4}(\partial_{\pm}F_{\pm})^{2}$ is a function related to the state of the quantum fields. 
Under a conformal coordinate transformation $y^{\pm}=y^{\pm}(x^{\pm})$ the energy-momentum tensor changes according to (we only consider the quantum correction)
\begin{equation}
\langle\Delta T^{(f)}_{\pm\pm}(y)\rangle=\left(\frac{d x^{\pm}}{d y^{\pm}}\right)^{2}\langle\Delta T^{(f)}_{\pm\pm}(x)\rangle.    
\end{equation}
For the conformal factor $\rho$, on the other hand, we have the fallowing transformation law,
\begin{equation}
\rho(y)=\rho(x)+\frac{1}{2}\ln\frac{d y^{+}}{d x^{+}}\frac{d y^{-}}{d x^{-}}. \end{equation}
Together, these give us the transformation law for $t_{\pm}$,
\begin{equation}\label{t_change}
t_{\pm}(y^{\pm})=\left(\frac{d x^{\pm}}{d y^{\pm}}\right)^{2}\left[t_{\pm}(x^{\pm})-\frac{1}{2}D_{x^{\pm}}[y^{\pm}]\right],    
\end{equation}
with Schwartz derivative defined by \begin{equation}
D_{x^{\pm}}[y^{\pm}]=\frac{(y^{\pm})'''}{(y^{\pm})'}-\frac{3}{2}\left(\frac{(y^{\pm})''}{(y^{\pm})'}\right)^{2},    
\end{equation}
where the derivatives are with respect to  $x^{\pm}$.

The vacuum state of quantum fields and the corresponding set of creation/anihilation operators depend on the reference frame, i.e. on the coordinate system. If we introduce normal ordering of the energy-momentum operator for one choice of the vacuum state $\vert 0; x\rangle$, say in  coordinate system $x^{\pm}$, the energy-momentum operator can be decomposed as
\begin{equation}\label{separation}
\hat{T}^{(f)}_{\pm\pm}(x^{\pm})=:\hat{T}^{(f)}_{\pm\pm}(x^{\pm}):+\langle 0;x\vert\hat{T}^{(f)}_{\pm\pm}(x^{\pm})\vert 0;x\rangle.     
\end{equation}
If we make a transition to another coordinate system $y^{\pm}$, the energy momentum tensor will not in general be normally ordered. The transformation law for the normally ordered part is given by
\begin{equation}\label{normal}
:\hat{T}^{(f)}_{\pm\pm}(y^{\pm}):=\left(\frac{d x^{\pm}}{d y^{\pm}}\right)^{2}\left[:\hat{T}^{(f)}_{\pm\pm}(x^{\pm}):+\frac{\hbar}{24\pi}D_{x^{\pm}}[y^{\pm}]\right].     
\end{equation}
Comparing (\ref{t_change}), (\ref{separation}) and (\ref{normal}), we can establish the following relationship,
\begin{equation}
\bra{0,x}:\hat{T}^{(f)}_{\pm\pm}(x^{\pm}):\ket{0,x}=-\epsilon t_{\pm}(x^{\pm}), \end{equation}
which provides an interpretation for the quantity $t_{\pm}$. In particular, since $\vert HH\rangle$ is the vacuum state in Kruskal coordinates, i.e. $\vert HH\rangle=\vert 0; x\rangle$, we have
\begin{equation}
\langle HH\vert : \hat{T}^{(f)}_{\pm\pm}(x^{\pm}):\vert HH\rangle=-\epsilon t_{\pm}(x^{\pm})=0. 
\end{equation}

Since we already have a factor of $\epsilon$ in (\ref{T_Kruskal}), we only need to consider $\partial_{\pm}\rho$ at the leading (classical) order, which is simply
\begin{align}
\partial_{\pm}\rho&=\frac{1}{2h}\partial_{\pm}h-\kappa\partial_{\pm}r_{*}=\frac{1}{4\kappa x^{\pm}}\left(\frac{r_{0}}{r^{2}}-\frac{1}{r_{0}}\right).
\end{align} 
This gives us
\begin{align}
\langle\Delta T^{(f)}_{\pm\pm}\rangle&=\epsilon\frac{(r-r_{0})^{2}}{r^{4}(x^{-})^{2}}\left(r^{2}+2r_{0}r+3r_{0}^{2}\right),\\
\langle\Delta T^{(f)}_{+-}\rangle&=\epsilon\frac{4r_{0}^{3}(r-r_{0})}{r^{4}x^{+}x^{-}}.
\end{align}
On the other hand, in Schwarzschild coordinates,
\begin{align}
\langle\Delta T^{(f)}_{rr}\rangle&=\frac{\partial x^{\mu}}{\partial r}\frac{\partial x^{\nu}}{\partial r}\langle\Delta T^{(f)}_{\mu\nu}\rangle\\
&=\left(\frac{\partial x^{+}}{\partial r}\right)^{2}\langle\Delta T^{(f)}_{++}\rangle+\left(\frac{\partial x^{-}}{\partial r}\right)^{2}\langle\Delta T^{(f)}_{--}\rangle\nonumber\\
&+2\frac{\partial x^{+}}{\partial r}\frac{\partial x^{-}}{\partial r}\langle\Delta T^{(f)}_{+-}\rangle=\frac{\epsilon}{2l^{2}}\left[\frac{1}{r_{0}^{2}}-\left(\frac{r_{0}}{r^{2}}\right)^{2}\right].\nonumber
\end{align}
Since $\partial_{r}h_{0}=\frac{r_{0}}{r^{2}}$, from (\ref{T_Sch_rr}) we can directly identify the constant $C$ as being equal to $\frac{1}{r_{0}}$, which is what we have expected. Now we have to determine the functions $m(r)$ and $\varphi(r)$.

Having computed the quantum corrections to the energy-momentum tensor, equations (\ref{diff_equations_m}) and (\ref{diff_equations_varphi}) become simple differential equations,
\begin{align}
\frac{dm}{dr}&=\frac{G_{\text{N}}}{\lambda^{2}r_{0}^{2}}\left[7\left(\frac{r_{0}}{r}\right)^{3}-\left(\frac{r_{0}}{r}\right)^{2}-\left(\frac{r_{0}}{r}\right)-1\right],\\
\frac{d\varphi}{dr}&=\frac{G_{\text{N}}}{\lambda^{2}r_{0}^{2}r}\left[3\left(\frac{r_{0}}{r}\right)^{2}+2\left(\frac{r_{0}}{r}\right)+1\right],
\end{align}
with solutions
\begin{align}
m(r)&=\frac{G_{\text{N}}}{\lambda^{2}r_{0}}\left[-\frac{7}{2}\left(\frac{r_{0}}{r}\right)^{2}+\frac{r_{0}}{r}+\ln{\frac{r_{0}}{r}}-\frac{r}{r_{0}}\right]+C_{1}, \\
\varphi(r)&=\frac{G_{\text{N}}}{\lambda^{2}r_{0}^{2}}\left[-\frac{3}{2}\left(\frac{r_{0}}{r}\right)^{2}-2\frac{r_{0}}{r}-\ln{\frac{r_{0}}{r}}\right]+C_{2}.\end{align}
We can also calculate the function $\alpha(r)$ defined in equation (\ref{r_zvezda}),
\begin{equation}
    \alpha(r)=\frac{G_{\text{N}}}{\lambda^{2}r^{2}_{0}}\left[\frac{5}{2}\frac{r}{r_{0}}+\left(1-\frac{r}{r_{0}}-\frac{r_{0}}{r-r_{0}}\right)\ln{\frac{r}{r_{0}}}\right].\label{r_zvezda1}
\end{equation}

Note that $\alpha(r_{H})=\frac{3G_{N}}{2\lambda^{2}r^{2}_{H}}$ does not diverge at the horizon, as we have already mentioned.
Functions $m(r)$ and $\varphi(r)$ are both defined up to an undetermined integration constant, $C_{1}$ and $C_{2}$, respectively. Note that $\varphi(r)$ diverges as $r\to+\infty$, which means that the metric does not appear to be asymptotically flat. A way to deal with this issue is to introduce a cut-off. Take $F(r):=\varphi(r)-C_{2}$ and define some characteristic dimension of space, $L$, such that $\varphi(r)=F(r)-F(L)$; now $\varphi\to0$ as $r\to L$. This fixes the asymptotic behavior of the constant $C_{2}$. We can choose $C_{2}=\frac{G_{N}}{\lambda^{2}r^{2}_{0}}\ln{\frac{r_{0}}{L}}$.
Function $m(r)$ also diverges, but this is not a problem since $\frac{m(r)}{r}$ is what actually appears in the metric.  We can also determine the constant $C_{1}$ by introducing another length scale, $l$, which should naturally be of order of Planck length, but this will not be important for further consideration. We can set $C_{1}=\frac{G}{\lambda^{2}r_{0}}\ln{\frac{l}{r_{0}}}$.

Finally, the position of the horizon and the surface gravity are given by
\begin{align}
r_{H}&=r_{0}-\epsilon m(r_{0})=r_{0}+\epsilon\left(\frac{7G_{\text{N}}}{2\lambda^{2}r_{0}}-C_{1}\right)\\
&=r_{0}+\frac{\epsilon G_{N}}{\lambda^{2}r_{0}}\left(\frac{7}{2}+\ln{\frac{r_{0}}{l}}\right),\\
\kappa&=\frac{1}{2r_{0}}\left[1+\epsilon\left(\varphi(r_{0})+\partial_{r}m(r_{0})+\frac{m(r_{0})}{r_{0}}\right)\right]\nonumber\\
&=\frac{1}{2r_{0}}\left[1+\epsilon\left(C_{2}+\frac{C_{1}}{r_{0}}-3\frac{G_{\text{N}}}{\lambda^{2}r_{0}^{2}}\right)\right]\\
&=\frac{1}{2r_{0}}\left[1-\frac{\epsilon G_{N}}{\lambda^{2}r^{2}_{0}}\left(3+\ln{\frac{L}{l}}\right)\right].
\end{align} 

We see that the horizon gets shifted due to quantum corrections. Also, since $T_{BH}=\kappa/2\pi$, the last equation implies that the black hole temperature acquires a quantum correction. 

\section{Page curve in DREH model}

Time-dependence of the fine-grained entropy of the Hawking radiation, as given by the QES formula (\ref{Island_formula}), essentially depends on the evolution of the spatial region $I\cup R$, where $I$ is to be determined by the variational method. We will consider two separate cases: one for which $I=\emptyset$ the whole time (no island), and the other when island appears at some point; the latter will give the unitary Page curve.

\subsection{No-island case}

\begin{figure}
    \begin{center}
    \includegraphics[width=7.5cm, height=4cm]{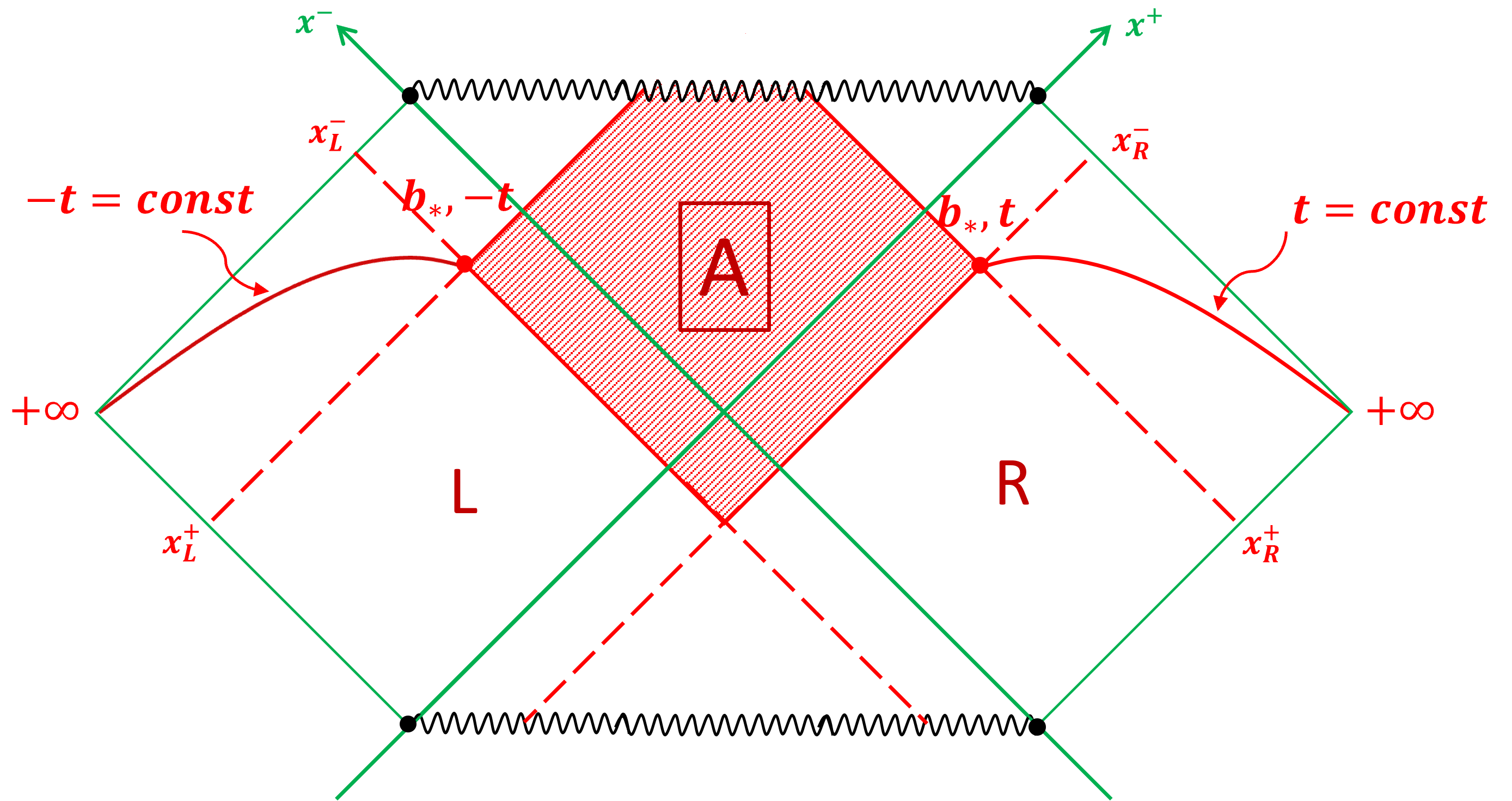}
    \caption{Position of the region ($A$) in which the inaccessible degrees of freedom live.}\label{sl1}
    \end{center}
\end{figure}

Let us first consider the situation with no island. In this case, we expect to reproduce the original Hawking's prediction. For an eternal black hole in Kruskal coordinates, the semi-classical entanglement entropy of the Hawking radiation can be computed as in \cite{QCGHS}, using the following formula (see Appendix A for some details), 
\begin{equation}
S_{matter}=\frac{1}{12}\ln{\frac{(x^{+}_{R}-x^{+}_{L})^{2}(x^{-}_{R}-x^{-}_{L})^{2}}{\delta^{4}e^{-2\rho_{R}}e^{-2\rho_{L}}}}.
\end{equation}
Labels $R/L$ correspond to the right/left asymptotically flat region of space-time, as shown in Figure \ref{sl1}. We can also work in $(t,r_{*})$ coordinates, or even $(t,r)$, since $r_{*}=r_{*}(r)$. Coordinates on the cutoff surface are then either $(t,b)$ or $(t,b_{*})$. Parameter $\delta$ is a UV cutoff. 
    
To recover the time dependence of the fine-grained entropy $S_{FG}$ along the cut-off surface, we make a transition to asymptotically flat coordinates that involves the modified surface gravity $\kappa$, 
\begin{align}
\kappa x^{+}_{R}&=e^{\kappa(t+b_{*})}\\
\kappa x^{-}_{R}&=-e^{-\kappa(t-b_{*})}\\
\kappa x^{+}_{L}&=-e^{\kappa(-t+b_{*})}\\
\kappa x^{-}_{L}&=e^{-\kappa(-t-b_{*})}.
\end{align}

In the DREH model we have $\rho=\frac{1}{2}\ln{h}+\epsilon\varphi-\kappa r_{*}$. Since $b$ is equal in the two asymptotically flat regions, it follows that $\rho_{R}=\rho_{L}$ and we get
\begin{equation}
S_{FG}=S_{matter}=\frac{1}{6}\ln{\frac{4h(b)\cosh^{2}{(\kappa t)}}{(\kappa\delta)^{2}e^{-2\epsilon\varphi(b)}}}.
\end{equation}

Regularization of the UV divergences is effected by demanding that $S_{FG}(0)=0$, i.e.
\begin{equation}
 S_{FG}=\frac{1}{3}\ln{\bigg{[}\cosh{(\kappa t)}\bigg{]}}.
\end{equation}

In the beginning of the evaporation (in the neighbourhood of $t=0$) the entropy behaves as $S_{FG}=\frac{1}{6}(\kappa t)^{2}$. However, we are more interested in the late-time limit, $\kappa t\gg1$. As in other dilaton gravity models \cite{GSST, Notes, HSS}, $S_{FG}$ grows linearly,
\begin{equation}
S_{FG}\approx\frac{1}{3}\kappa t-\frac{1}{3}\ln{2}.
\end{equation}
Thus we conclude 
that the fine-grained entropy of the radiation increases monotonically, which is in agreement with the original Hawking's result, but in contrast with the RST and BPP model, the DREH model predicts that the surface gravity acquires quantum correction.

\subsection{Island case}

In the presence of an island, the relevant region consists of two disjoint parts and we have to use the following formula (see Appendix A for some details),
\begin{equation}
S_{matter}=\frac{1}{6}\ln{\frac{d^{2}_{12}d^{2}_{23}d^{2}_{14}d^{2}_{34}}{\delta^{4}d^{2}_{24}d^{2}_{13}e^{-\rho_{1}}e^{-\rho_{2}}e^{-\rho_{3}}e^{-\rho_{4}}}},
\end{equation}
with
\begin{equation}
d^{2}_{ij}=(x^{+}_{i}-x^{+}_{j})(x^{-}_{i}-x^{-}_{j}).    
\end{equation}

The position of the island is given by $(t',a_{*})$ in the asymptotically flat region on the right. In the left region the corresponding point is symmetrical with respect to the vertical axis. Again, we vary with respect to $t'$ and $a_{*}$.
    
\begin{figure}
    \begin{center}
    \includegraphics[width=7.5cm, height=4cm]{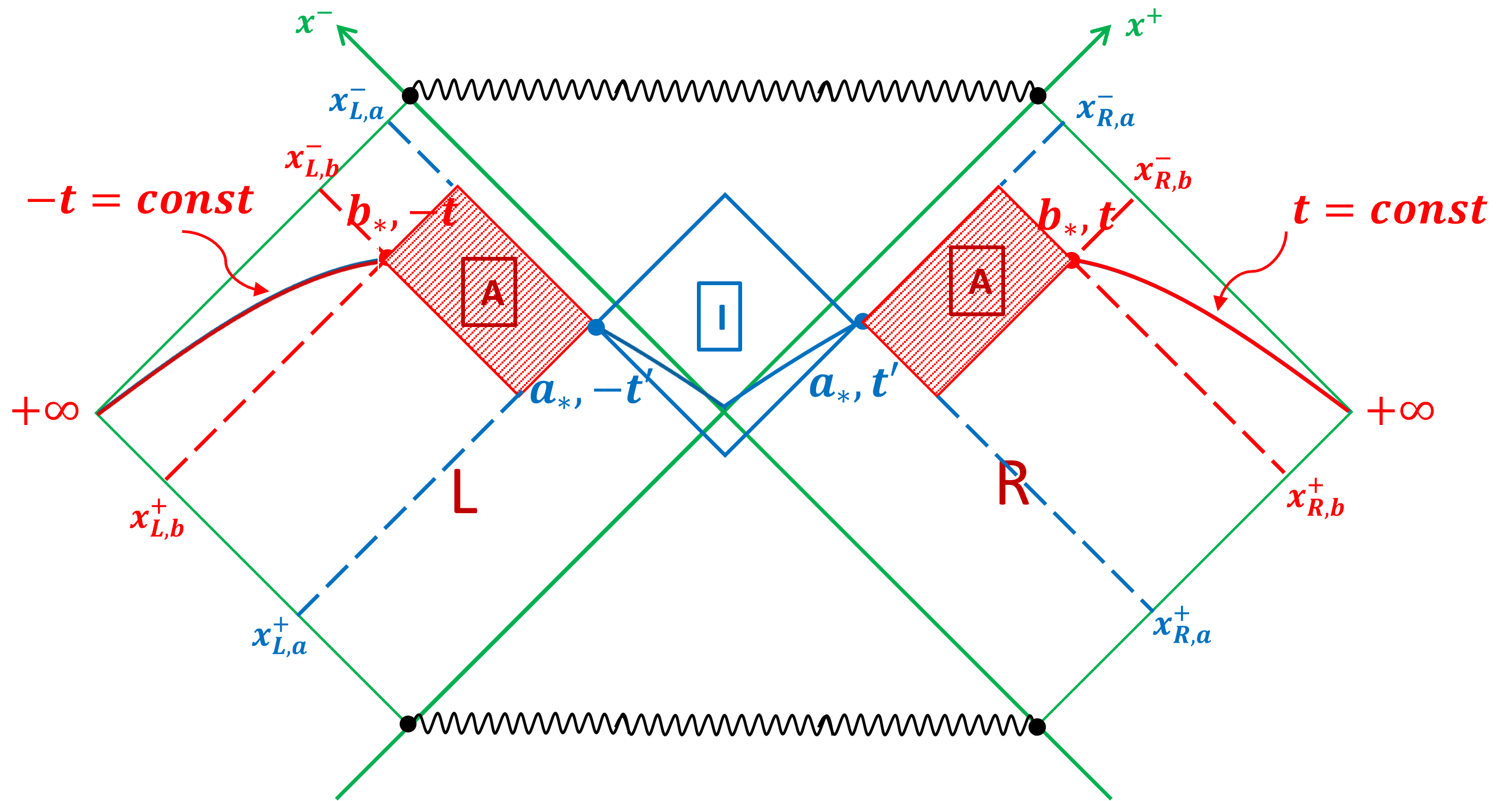}
    \caption{Position of the island ($I$) and the disconnected region ($A$) in witch the inaccessible degrees of freedom live.}\label{sl2}
    \end{center}
\end{figure}
  
In asymptotically flat coordinates, the position of the region $A$ on Figure 2 is given by
\begin{align}
\kappa x^{+}_{Rb}&=e^{\kappa(t+b_{*})},\hspace{0.5cm}\kappa x^{-}_{Rb}=-e^{-\kappa(t-b_{*})}\label{t1},\\
\kappa x^{+}_{Lb}&=-e^{\kappa(-t+b_{*})},\hspace{0.5cm}\kappa x^{-}_{Lb}=e^{-\kappa(-t-b_{*})}\label{t2},\\
\kappa x^{+}_{Ra}&=e^{\kappa(t'+a_{*})},\hspace{0.5cm}
\kappa x^{-}_{Ra}=-e^{-\kappa(t'-a_{*})}\label{t3},\\
\kappa x^{+}_{La}&=-e^{\kappa(-t'+a_{*})},\hspace{0.5cm}
\kappa x^{-}_{La}=e^{-\kappa(-t'-a_{*})}\label{t4}.
\end{align}

Now we consider the late time stage of the evaporation process. As in the case with no island, we have that $\frac{d^{2}_{23}d^{2}_{14}}{d^{2}_{24}d^{2}_{13}}\to1$ (see Appendix A), which means that the entropy formula comes down to
\begin{equation}
    S_{matter}=\frac{1}{6}\ln{\left(\frac{d^{4}_{12}}{\delta^{4}}e^{2\rho_{a}}e^{2\rho_{b}}\right)}.
\end{equation}

After making the substitutions for $\rho$ and $d_{12}$, the above formula becomes 
\begin{widetext}
\begin{equation}
S_{matter}=\frac{1}{6}\ln{\frac{h(a)h(b)\left(e^{\kappa(t+b_{*})}-e^{\kappa(t'+a_{*})}\right)^{2}\left(-e^{-\kappa(t-b_{*})}+e^{-\kappa(t'-a_{*})}\right)^{2}}{(\kappa\delta)^{4}e^{2\kappa(b_{*}+a_{*})}e^{-2\varepsilon(\varphi(a)+\varphi(b))}}}.
\end{equation}
\end{widetext}

The surface term does not depend on $t'$, so we can immediately perform extremization over $t'$. For late times it holds that $t'=t$ as in \cite{BPP}. Now the formula for $S_{matter}$ becomes
\begin{align}
S_{matter}=\frac{1}{3}(\rho_{a}+\rho_{b})&+\frac{2}{3}\ln{\left(e^{\kappa b_{*}}-e^{\kappa a_{*}}\right)}\\\nonumber-\frac{2}{3}\ln{\left(\kappa\delta\right)}.
\end{align}
The surface term of the generalized entropy reads
\begin{equation}
\frac{A[I]}{4G^{(4)}_{\text{N}}}=2\frac{4\pi \lambda^{2}a^{2}}{4G_{\text{N}}\hbar}.
\end{equation}
This is exactly the Bekenstein-Hawking entropy. Note that we have the combination $\lambda^{2}/G_{\text{N}}$, since the formula holds in four dimensions. The factor of two comes form having two asymptotically flat regions. 

Therefore, the generalized entropy is 
\begin{align}
S_{gen}=\frac{2\pi \lambda^{2}a^{2}}{G_{\text{N}}\hbar}+\frac{1}{3}\left(\rho_{a}+\rho_{b}\right)&+\frac{2}{3}\ln{\left(e^{\kappa b_{*}}-e^{\kappa a_{*}}\right)}\nonumber\\
&-\frac{2}{3}\ln{(\kappa\delta)}.
\end{align}

Now, we extremize in terms of the position $a_{*}$,
\begin{equation}
\partial_{a_{*}}S_{gen}=\left[\frac{4\pi\lambda^{2} a}{G_{\text{N}}\hbar}+\frac{1}{3}\frac{d\rho(a)}{dr}\right]\frac{da}{da_{*}}-\frac{2}{3}\frac{\kappa}{e^{\kappa(b_{*}-a_{*})}-1}=0,
\end{equation}
which gives us
\begin{equation}\label{extremal}
\left[a+\frac{4\epsilon G_{N}}{\lambda^{2}}\frac{d\rho(a)}{dr}\right]h(a)e^{\epsilon\varphi(a)}=\frac{8\epsilon G_{N}}{\lambda^{2}}\frac{\kappa}{e^{\kappa(b_{*}-a_{*})}-1}.
\end{equation}

We expect to find QES near the horizon. In the case of eternal black hole it should be outside the horizon. So we write: $a=r_{H}+x$, where $r_{H}$ is a quantum corrected position of the horizon, and $x\ll r_{H}$. Then we can expand left hand side of the equation (\ref{extremal}) with respect to small $x$, up to the first order. We get
\begin{align}
&\rm{LHS}\approx\left[r_{H}+\frac{4\epsilon G_{N}}{\lambda^{2}}\frac{d\rho}{dr}\bigg{|}_{H}\right]h(r_{H})e^{\epsilon\varphi(r_{H})}\nonumber\\
&+\left[r_{H}+\frac{4\epsilon G_{N}}{\lambda^{2}}\frac{d\rho}{dr}\bigg{|}_{H}\right]\left(\frac{dh}{dr}e^{\epsilon\varphi(r_{H})}\right)\bigg{|}_{H}x\label{leva}
\end{align}
The first term vanishes since $h(r_{H})=0$ by a definition of a horizon position. While, in the second term, we have: $\left(\frac{dh}{dr}e^{\epsilon\varphi}\right)\bigg{|}_{H}=2\kappa$, by a definition of the surface gravity. Now, we consider the right hand side of equation (\ref{extremal}) and expand $e^{2\kappa a_{*}}$ near the horizon,
\begin{align}
    e^{2\kappa a_{*}}&\approx e^{2\kappa r_{*}(r_{H})}+2\kappa\left(e^{2\kappa r_{*}}\frac{dr_{*}}{dr}\right)\bigg{|}_{H}x\nonumber\\
    &=2\kappa xe^{1+\epsilon(\alpha(r_{H})-\varphi(r_{H}))}\lim_{r\to r_{H}}\left(\frac{\frac{r}{r_{H}}-1}{h(r)}\right)\nonumber\\
    &=2\kappa xe^{1+\epsilon\alpha(r_{H})}\frac{e^{-\epsilon\varphi(r_{H})}}{r_{H}h'(r_{H})}=\frac{x}{r_{H}}e^{1+\epsilon\alpha(r_{H})}.
\end{align}
The first term in the first line vanishes since $r_{*}(r_{H})\to\infty$ and in the second line we used formula (\ref{r_zvezda}). The right hand side of the equation (\ref{extremal}) now becomes
\begin{align}
    &\rm{RHS}\approx\frac{8\epsilon\kappa G_{N}}{\lambda^{2}}e^{-\kappa b_{*}}e^{\kappa a_{*}}\left(1+e^{-\kappa b_{*}}e^{\kappa a_{*}}\right)\label{desna}\\
    &=\frac{8\epsilon\kappa G_{N}}{\lambda^{2}}\left[\sqrt{\frac{x}{r_{H}}}e^{\frac{1}{2}(1+\epsilon\alpha(r_{H}))-\kappa b_{*}}+\frac{x}{r_{H}}e^{1+\epsilon\alpha(r_{H})-2\kappa b_{*}}\right].\nonumber
\end{align}
Using the equations (\ref{leva}) i (\ref{desna}) and solving for $x$, we get
\begin{equation}
    x=\frac{1}{r_{H}}\frac{\left(\frac{4\epsilon G_{N}}{\lambda^{2}r_{H}}\right)^{2}e^{1-2\kappa b_{*}+\epsilon\alpha(r_{H})}}{\left[1+\frac{4\epsilon G_{N}}{\lambda^{2}r_{H}}\left(\frac{d\rho}{dr}\bigg{|}_{H}-\frac{1}{r_{H}}e^{1-2\kappa b_{*}+\epsilon\alpha(r)}\right)\right]^{2}}.
\end{equation}


Using the equations (\ref{2}) and (\ref{r_zvezda1}), and remembering that $\epsilon=\frac{\hbar}{48\pi}$, we get the position of the island,
\begin{widetext}
\begin{equation}\label{Islan_position}
    a=r_{H}+\frac{1}{r_{H}}\left(\frac{\hbar G_{N}}{12\pi\lambda^{2}r_{H}}\right)^{2}e^{1-2\kappa b_{*}}\left[1+\frac{\hbar G_{N}}{6\pi\lambda^{2}r^{2}_{H}}\left(\frac{19}{16}+e^{1-2\kappa b_{*}}\right)\right].
\end{equation}
\end{widetext}

Note that the result(\ref{Islan_position}) exactly coincides with \cite{HIM} in the leading order of $\epsilon$ expansion. The main difference is that we have also included back-reaction of the radiation, which means that quantum corrections do not spoil the fact that an island appears close to the (quantum corrected) horizon.

Now we can determine the behaviour of the fine-grained entropy. At late times, and expanding around $r_{H}$, we get
\begin{equation}
    S_{FG}=S_{gen}(r_{H})-\frac{\hbar G_{N}}{36\pi\lambda^{2}r_{H}^{2}}e^{1-2\kappa b_{*}}+\mathcal{O}(\hbar^{2})\ .
\end{equation}
The second term can be neglected since it is of order $\mathcal{O}(\hbar)$, and we are left with
\begin{align}
    S_{FG}&=\frac{2\pi \lambda^{2}r^{2}_{H}}{G_{\text{N}}\hbar}+\frac{1}{3}\left(\rho_{H}+\rho_{b}\right)+\frac{2}{3}\ln{\frac{e^{\kappa b_{*}}}{\kappa\delta}}\\
    &=2\left[\frac{\pi \lambda^{2}r^{2}_{H}}{G_{\text{N}}\hbar}+\frac{1}{12}\ln{\frac{e^{4\kappa b_{*}}}{(\kappa\delta)^{4}e^{-2\rho_{H}}e^{-2\rho_{b}}}}\right].
\end{align}
Comparing with equations (\ref{wald1}), (\ref{wald2}) and (\ref{wald3}), we have $S_{gen}=2S_{BH}$, as expected. The fine-grained entropy of the radiation is therefore
\begin{equation}
S_{FG}=\min\bigg{\{}\frac{1}{3}\kappa t,2S_{BH}\bigg{\}}.
\end{equation}

To determine the Page time, we use the classical value of entropy since the quantum correction is dwarfed in the large $M$ limit. The Page time $t_{P}$ is determined by
\begin{equation}
\frac{1}{3}\kappa ct_{P}=\frac{2\pi c^{3} \lambda^{2}r_{0}^{2}}{G_{\text{N}}\hbar},
\end{equation}
and so
\begin{equation}
t_{P}=\frac{96\pi M^{3}G_{\text{N}}^{2}}{\hbar\lambda^{4}c^{4}}.\end{equation}

\section{Conclusion} 

We studied a model of two-dimensional dilaton gravity related to four-dimensional Einstein-Hilbert action by dimensional reduction. The Page curve is successfully reproduced for an eternal Schwarzschild black hole. The fine-grained entropy of the radiation grows monotonically until it reaches the course-grained Bekenstein-Hawking limit; after that, it remains equal to $2S_{BH}$. An important aspect of the DREH model is that the Hawking temperature acquires a quantum correction, which is not the case, for example, in the related BPP model. Further investigation might include analysis of a more realistic scenario involving an evaporating black hole or
dimensional reduction of the electromagnetic field coupled to gravity and the analysis of quantum corrections to the Reissner-Nordstrom solution or AdS-Schwawzschild black hole.
Another important opservation is that the Hawking's result for the generalized entropy, which corresponds to the no-island case, can also be reproduced using the Wald's formula that treats entropy as a Noether charge (See appendix B). It would be interesting to consider a modification of the Wald's entropy formula to include the island scenario.

{\bf Acknowledgments}
A.G., D.G. and V.R. acknowledge the funding provided by the Faculty of Physics, University of Belgrade, through the grant by the Ministry of Education, Science, and Technological Development of the Republic of Serbia.

\appendix
\setcounter{figure}{0}
\renewcommand{\thefigure}{A\arabic{figure}}
\section*{APPENDIX}
\section{Entanglement entropy formula}

Here we give a brief review of the derivation of the QFT entanglement entropy formula (denoted by $S_{matter}$ in the equation (\ref{Island_formula})), closely following the account of \cite{QCGHS}. For that matter, let us first recall the Unruh's result for the entropy of a uniformly accelerating observer in Minkowski spacetime, also known as the Rindler observer. If the world-line of a Rindler observer belongs to the right Rindler wedge, the observer cannot access the degrees of freedom which live in the causally disconnected left Rindler wedge. After tracing out the inaccessible degrees of freedom, the Minkowski vacuum, which is a pure vacuum state for all inertial observers, reduces to a thermal state for a Rindler observer. The von Neumann entropy of this reduced state is known as entanglement entropy, and it is given by
\begin{equation}
S_{ent}=\frac{1}{12}\ln{\frac{X^{+}_{max}X^{-}_{max}}{\delta^{2}}},\label{ent1}
\end{equation}
where $X^{\pm}_{max}$ stand for IC cutoffs in the light cone directions while $\delta$ represents a UV cutoff. 

\begin{figure}[h]
    \begin{center}
    \includegraphics[width=5cm, height=5cm]{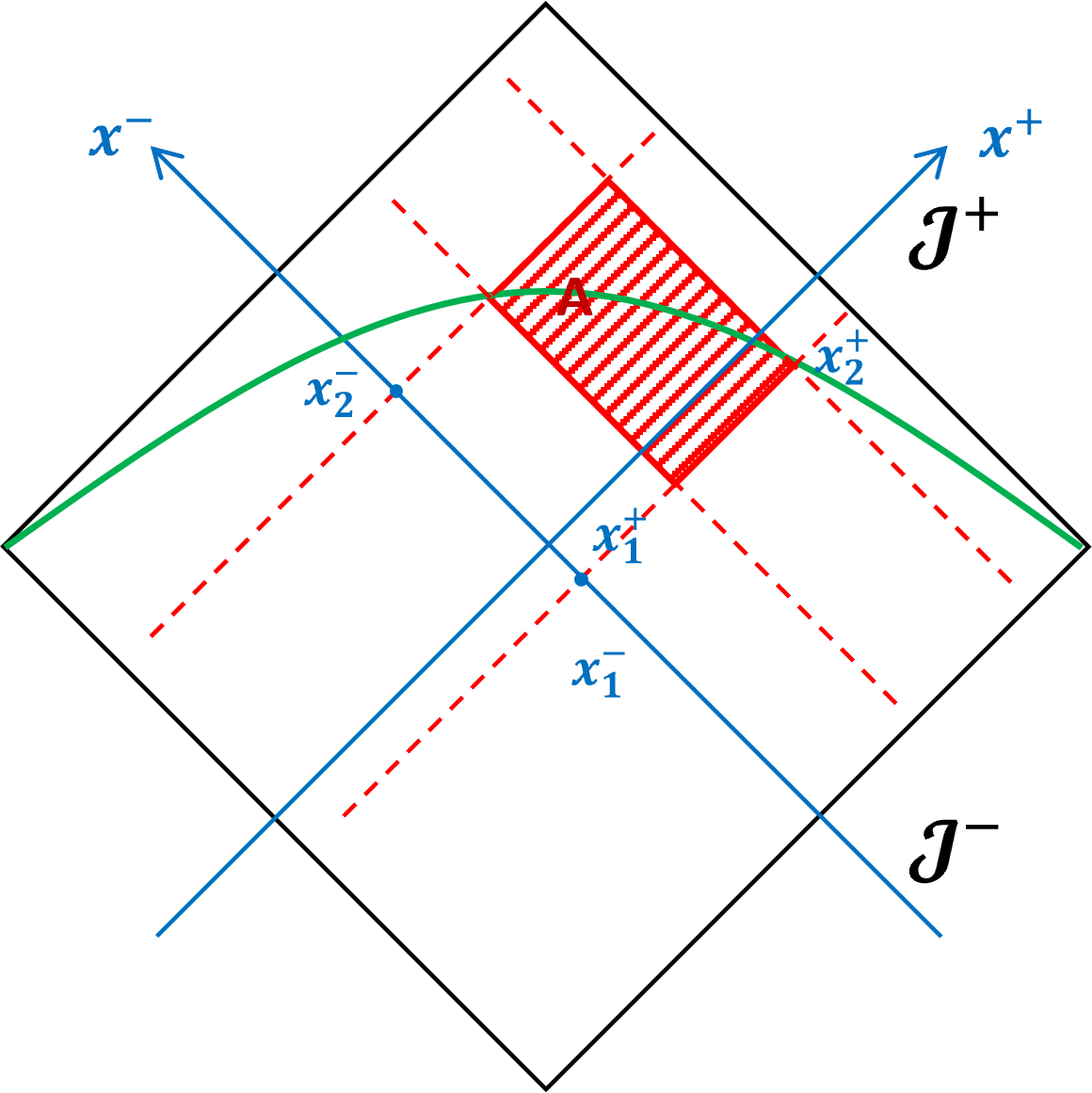}
    \caption{Position of the region ($A$) in witch the inaccessible degrees of freedom live.}\label{sl3}
    \end{center}
\end{figure}

The next step is to generalize the previous formula to include 
an arbitrary inaccessible region of Minkowski spacetime, see figure \ref{sl3}. As shown in \cite{QCGHS}, the appropriate generalization is

\begin{equation}\label{S_ent}
    S_{ent}=\frac{1}{12}\ln{\frac{(x^{+}_{2}-x^{+}_{1})^{2}(x^{-}_{2}-x^{-}_{1})^{2}}{\delta^{4}}}.
\end{equation}
The curved spacetime version of the entanglement entropy formula can be obtained in two steps. 
First, one can find how formula (\ref{S_ent}) looks in some other flat spacetime coordinates $y^{\pm}=y^{\pm}(x^{\pm})$, and then conclude that the same formula holds in curved spacetime as well. Since the calculation is the same as in Minkowski coordinates, one gets

\begin{equation}
    S_{ent}=\frac{1}{12}\ln{\frac{(y^{+}_{2}-y^{+}_{1})^{2}(y^{-}_{2}-y^{-}_{1})^{2}}{\hat{\delta}^{4}}},
\end{equation}
where $\hat{\delta}$ is a UV cutoff in $y$-coordinates. However, this is not a proper UV cutoff since it is varies from one point of spacetime to another. It transforms as a length, so it is easy to transform it back to globally flat Minkowski coordinates. In conformal gauge, $ds^{2}=-e^{2\rho}dy^{+}dy^{-}$, this yields
\begin{equation}
S_{ent}=\frac{1}{12}\ln{\frac{(y^{+}_{2}-y^{+}_{1})^{2}(y^{-}_{2}-y^{-}_{1})^{2}}{\delta^{4}e^{-2\rho_{1}}e^{-2\rho_{2}}}}.\label{ent4}
\end{equation}
Since one can choose to define $\delta$ in locally flat coordinates, it is easy to conclude that formula (\ref{ent4}) holds in curved spacetime as well. 

\begin{figure}[h]
    \begin{center}
    \includegraphics[width=5.5cm, height=4.6cm]{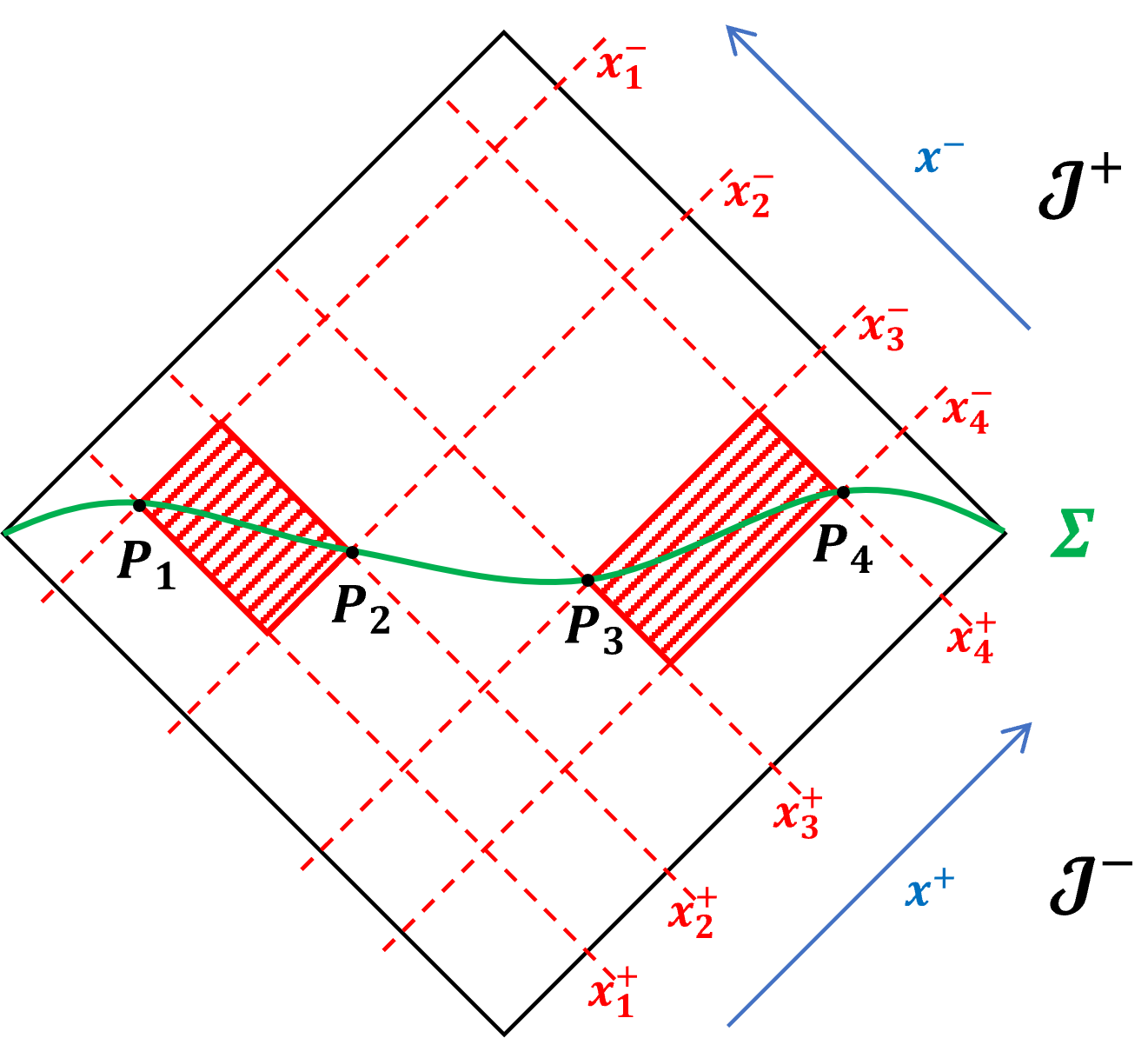}
    \caption{Position of the disconnected region ($A$) in witch the inaccessible degrees of freedom live.}\label{sl4}
    \end{center}
\end{figure}

In the case when an island is present, there exist two disjoint regions of spacetime (see figure \ref{sl4}) that one has to take into account. Using a similar reasoning as in \cite{QCGHS} one can obtain the following formula for entanglement entropy
\begin{equation}
S_{ent}=\frac{1}{6}\ln{\frac{d^{2}_{12}d^{2}_{23}d^{2}_{14}d^{2}_{34}}{\delta^{4}d^{2}_{24}d^{2}_{13}e^{-\rho_{1}}e^{-\rho_{2}}e^{-\rho_{3}}e^{-\rho_{4}}}},\label{ent3}
\end{equation}
where $d^{2}_{ij}=(x^{+}_{i}-x^{+}_{j})(x^{-}_{i}-x^{-}_{j})$ is Minkowski distance between two endpoints of the two inaccessible regions. The formula (\ref{ent3}) holds in curved spacetime as well. 

In the late time limit, one can show that 
\begin{equation}
\frac{d^{2}_{23}d^{2}_{14}}{d^{2}_{24}d^{2}_{13}}\to1.    
\end{equation}
Using the coordinate transformations (\ref{t1}), (\ref{t2}), (\ref{t3}) and (\ref{t4}) this yields
\begin{widetext}
\begin{equation}
    \frac{d^{2}_{23}d^{2}_{14}}{d^{2}_{24}d^{2}_{13}}=16e^{2\kappa(b_{*}-a_{*})}\frac{\cosh^{2}{(\kappa t)}\cosh^{2}{(\kappa t')}}{\left(e^{\kappa(t+b_{*})}+e^{-\kappa(t'+a_{*})}\right)^{2}\left(e^{\kappa(b_{*}-t)}+e^{\kappa(t'-a_{*})}\right)^{2}}\approx1;\hspace{2mm}t,t'\to\infty.
\end{equation}
\end{widetext}

It is also easy to show that $d_{12}=d_{34}$, $\rho_{1}=\rho_{4}$ and $\rho_{2}=\rho_{3}$ in this case. Using these results, equation (\ref{ent3}) reduces to

\begin{equation}
    S_{ent}=\frac{1}{6}\ln{\left(\frac{d^{4}_{12}}{\delta^{4}}e^{2\rho_{1}}e^{2\rho_{2}}\right)}.
\end{equation}

\section{Wald's entropy}

In Wald's formulation, the entropy of a black hole is understood as a charge at the horizon. It is given by

\begin{equation}S_{Wald}=-2\pi\int_{H} dA\frac{\partial\mathcal{L}}{\partial\mathcal{R}_{\mu\nu\rho\sigma}}\varepsilon_{\mu\nu}\varepsilon_{\rho\sigma},
\end{equation}
where $\varepsilon_{\mu\nu}$ is a binormal satisfying $\varepsilon_{\mu\nu}\varepsilon^{\mu\nu}=-2$, $dA$ the infinitesimal area element of the event horizon and $\mathcal{L}$ the Lagrangian density defining the theory. A direct calculation in the case of (\ref{dejstvo}) yields

\begin{equation}
S_{Wald}=\frac{\pi\lambda^{2} r_{H}^{2}}{G_{\text{N}}\hbar}-\frac{1}{12}\psi(r_{H}).    \label{wald1}
\end{equation}
Solving the equation for the auxiliary field $\psi$ in Kruskal coordinates yields $\psi=-2\rho+F_{+}(x^{+})+F_{-}(x^{-})$. Functions $F_{\pm}(x^{\pm})$ can be calculating using the definition of the HH state, namely  $t_{\pm}(x^{\pm})=0$. This gives us
\begin{equation}
    \frac{1}{2}\partial_{\pm}^{2}F_{\pm}-\frac{1}{4}(\partial_{\pm}F_{\pm})^{2}=0.
\end{equation}
Solving the equation yields
\begin{equation}
    F_{\pm}(x^{\pm})=-2\ln{\frac{x^{\pm}+c_{\pm}}{\delta}}+d_{\pm},
\end{equation}
where $c_{\pm}$ and $d_{\pm}$ are integration constants. We want to interpret constants $c_{\pm}$ as positions in space-time. To make $\psi$ invariant under the local Lorentz transformations, we need to choose $d_{+}+d_{-}=-2\rho_{c}$. After implementing this condition we get
\begin{equation}
\psi(r)=-\ln\frac{(x^{+}+c_{+})^{2}(x^{-}+c_{-})^{2}}{\delta^{4}e^{-2\rho}e^{-2\rho_{c}}},\label{wald2}
\end{equation}
where $c_{\pm}$ should be chosen at the cut-off surface that defines a region of space that belongs to the black hole (see Figures \ref{sl1} and \ref{sl2}). This means that $\kappa^{2}c_{+}c_{-}=-e^{2\kappa b_{*}}$. To compute Wald's entropy we have to evaluate (\ref{wald2}) at the horizon where $x^{+}_{H}=x^{-}_{H}=0$. The Wald's entropy can thus be seen as quantum corrected Bekenstein-Hawking formula,
\begin{equation}
    S_{Wald}=\frac{\pi\lambda^{2} r_{H}^{2}}{G_{\text{N}}\hbar}+\frac{1}{12}\ln{\frac{(e^{2\kappa b_{*}})^{2}}{(\kappa\delta)^{4}e^{-2\rho_{H}}e^{-2\rho_{b}}}}.\label{wald3}
\end{equation}


\begin{thebibliography}{99}

\bibitem{Hawking1} S.~W.~Hawking,
``Breakdown of Predictability in Gravitational Collapse,''
Phys. Rev. D \textbf{14}, 2460-2473 (1976).


\bibitem{Bekenstein1} J.~D.~Bekenstein,
``Black holes and the second law,''
Lett. Nuovo Cim. \textbf{4}, 737-740 (1972).

\bibitem{Bekenstein2} J.~D.~Bekenstein,
``Black holes and entropy,''
Phys. Rev. D \textbf{7}, 2333-2346 (1973).

\bibitem{Hawking2} S.~W.~Hawking,
``Black Holes and Thermodynamics,''
Phys. Rev. D \textbf{13}, 191-197 (1976).

\bibitem{Hawking3} S.~W.~Hawking,
``Particle Creation by Black Holes,''
Commun. Math. Phys. \textbf{43}, 199-220 (1975).

\bibitem{JT1} R.~Jackiw,
``Lower Dimensional Gravity,''
Nucl. Phys. B \textbf{252}, 343-356 (1985).

\bibitem{JT2} C.~Teitelboim,
``Gravitation and Hamiltonian Structure in Two Space-Time Dimensions,''
Phys. Lett. B \textbf{126}, 41-45 (1983). 

\bibitem{CGHS} C.~G.~Callan, Jr., S.~B.~Giddings, J.~A.~Harvey and A.~Strominger,
``Evanescent black holes,''
Phys. Rev. D \textbf{45}, no.4, R1005 (1992).

\bibitem{RST1} J.~G.~Russo, L.~Susskind and L.~Thorlacius,
``Cosmic censorship in two-dimensional gravity,''
Phys. Rev. D \textbf{47}, 533-539 (1993).
 
\bibitem{RST2} J.~G.~Russo, L.~Susskind and L.~Thorlacius,
``The Endpoint of Hawking radiation,''
Phys. Rev. D \textbf{46}, 3444-3449 (1992). 

\bibitem{BPP} S.~Bose, L.~Parker and Y.~Peleg,
``Semiinfinite throat as the end state geometry of two-dimensional black hole evaporation,''
Phys. Rev. D \textbf{52}, 3512-3517 (1995).

\bibitem{QCGHS} T.~M.~Fiola, J.~Preskill, A.~Strominger and S.~P.~Trivedi,``Black hole thermodynamics and information loss in two-dimensions,'' Phys. Rev. D \textbf{50}, 3987-4014 (1994).

\bibitem{Fabri} Fabbri, Alessandro, and Jos Navarro-Salas. Modeling black hole evaporation. World Scientific, 2005.

\bibitem{DWV} D.~Grumiller, W.~Kummer and D.~V.~Vassilevich,
``Dilaton gravity in two-dimensions,''
Phys. Rept. \textbf{369}, 327-430 (2002).

\bibitem{Page1}
D.~N.~Page,
``Information in black hole radiation,''
Phys. Rev. Lett. \textbf{71}, 3743-3746 (1993).
 
\bibitem{Page2} 
D.~N.~Page,
``Time Dependence of Hawking Radiation Entropy,''
JCAP \textbf{09}, 028 (2013).
 
\bibitem{Penington} G.~Penington,
``Entanglement Wedge Reconstruction and the Information Paradox,''
JHEP \textbf{09}, 002 (2020). 
 
\bibitem{AEMM} A.~Almheiri, N.~Engelhardt, D.~Marolf and H.~Maxfield,
``The entropy of bulk quantum fields and the entanglement wedge of an evaporating black hole,''
JHEP \textbf{12}, 063 (2019).

\bibitem{AMMZ} A.~Almheiri, R.~Mahajan, J.~Maldacena and Y.~Zhao,
``The Page curve of Hawking radiation from semiclassical geometry,''
JHEP \textbf{03}, 149 (2020).

\bibitem{AMM} A.~Almheiri, R.~Mahajan and J.~Maldacena,
``Islands outside the horizon,''
[arXiv:1910.11077 [hep-th]].

\bibitem{RT} S.~Ryu and T.~Takayanagi,
``Holographic derivation of entanglement entropy from AdS/CFT,''
Phys. Rev. Lett. \textbf{96}, 181602 (2006).

\bibitem{HRT} V.~E.~Hubeny, M.~Rangamani and T.~Takayanagi,
``A Covariant holographic entanglement entropy proposal,''
JHEP \textbf{07}, 062 (2007).

\bibitem{QES} N.~Engelhardt and A.~C.~Wall,
``Quantum Extremal Surfaces: Holographic Entanglement Entropy beyond the Classical Regime,''
JHEP \textbf{01}, 073 (2015).
 
\bibitem{CW} C.~G.~Callan, Jr. and F.~Wilczek,
``On geometric entropy,''
Phys. Lett. B \textbf{333}, 55-61 (1994). 

\bibitem{HLW} C.~Holzhey, F.~Larsen and F.~Wilczek,
``Geometric and renormalized entropy in conformal field theory,''
Nucl. Phys. B \textbf{424}, 443-467 (1994).

\bibitem{CC} P.~Calabrese and J.~Cardy,
``Entanglement entropy and conformal field theory,''
J. Phys. A \textbf{42}, 504005 (2009).

\bibitem{PSSY} G.~Penington, S.~H.~Shenker, D.~Stanford and Z.~Yang,
``Replica wormholes and the black hole interior,''
[arXiv:1911.11977 [hep-th]]. 
 
\bibitem{AHMST} A.~Almheiri, T.~Hartman, J.~Maldacena, E.~Shaghoulian and A.~Tajdini,
``Replica Wormholes and the Entropy of Hawking Radiation,''
JHEP \textbf{05}, 013 (2020). 
 
\bibitem{CFHMR} H.~Z.~Chen, Z.~Fisher, J.~Hernandez, R.~C.~Myers and S.~M.~Ruan,
``Information Flow in Black Hole Evaporation,''
JHEP \textbf{03}, 152 (2020). 
 
\bibitem{Chen} Y.~Chen,
``Pulling Out the Island with Modular Flow,''
JHEP \textbf{03}, 033 (2020). 

\bibitem{AEPU} C.~Akers, N.~Engelhardt, G.~Penington and M.~Usatyuk,
``Quantum Maximin Surfaces,''
JHEP \textbf{08}, 140 (2020).

\bibitem{LV} H.~Liu and S.~Vardhan,
``A dynamical mechanism for the Page curve from quantum chaos,''
JHEP \textbf{03}, 088 (2021).
 
\bibitem{MM} D.~Marolf and H.~Maxfield,
``Transcending the ensemble: baby universes, spacetime wormholes, and the order and disorder of black hole information,''
JHEP \textbf{08}, 044 (2020). 

\bibitem{BKPSU} V.~Balasubramanian, A.~Kar, O.~Parrikar, G.~S\'arosi and T.~Ugajin,
``Geometric secret sharing in a model of Hawking radiation,''
JHEP \textbf{01}, 177 (2021).
 
\bibitem{Bhattacharya} A.~Bhattacharya,
``Multipartite purification, multiboundary wormholes, and islands in $AdS_3/CFT_2$,''
Phys. Rev. D \textbf{102}, no.4, 046013 (2020). 
\bibitem{Verlinde} H.~Verlinde,
``ER = EPR revisited: On the Entropy of an Einstein-Rosen Bridge,''
[arXiv:2003.13117 [hep-th]].

\bibitem{CQZ} Y.~Chen, X.~L.~Qi and P.~Zhang,
``Replica wormhole and information retrieval in the SYK model coupled to Majorana chains,''
JHEP \textbf{06}, 121 (2020).

\bibitem{GSST} F.~F.~Gautason, L.~Schneiderbauer, W.~Sybesma and L.~Thorlacius,
``Page Curve for an Evaporating Black Hole,''
JHEP \textbf{05}, 091 (2020). 

\bibitem{APS}  L.~Anderson, O.~Parrikar and R.~M.~Soni,
``Islands with gravitating baths: towards ER = EPR,''
JHEP \textbf{21}, 226 (2020). 
 
\bibitem{GHT} K.~Goto, T.~Hartman and A.~Tajdini,
``Replica wormholes for an evaporating 2D black hole,''
JHEP \textbf{04}, 289 (2021). 
 
\bibitem{AI} T.~Anegawa and N.~Iizuka,
``Notes on islands in asymptotically flat 2d dilaton black holes,''
JHEP \textbf{07}, 036 (2020). 

\bibitem{AMS} A.~Almheiri, R.~Mahajan and J.~E.~Santos,
``Entanglement islands in higher dimensions,''
SciPost Phys. \textbf{9}, no.1, 001 (2020).  

\bibitem{HIM} K.~Hashimoto, N.~Iizuka and Y.~Matsuo,
``Islands in Schwarzschild black holes,''
JHEP \textbf{06}, 085 (2020).

\bibitem{WLW} X.~Wang, R.~Li and J.~Wang,
``Islands and Page curves of Reissner-Nordstr\"om black holes,''
JHEP \textbf{04}, 103 (2021). 

\bibitem{Q} M.~H.~Yu and X.~H.~Ge,
``Islands and Page curves in charged dilaton black holes,''
Eur. Phys. J. C \textbf{82}, no.1, 14 (2022).

\bibitem{MVM} M.~Buric, V.~Radovanovic and A.~R.~Mikovic,
``One loop correction for Schwarzschild black hole via 2-D dilaton gravity,''
Phys. Rev. D \textbf{59}, 084002 (1999).
 
\bibitem{Polyakov} A.~M.~Polyakov,
``Quantum Geometry of Bosonic Strings,''
Phys. Lett. B \textbf{103}, 207-210 (1981). 

\bibitem{Frolov} V.P.~Frolov, W.~Israel and S.N.~ Solodukhin, ``One-loop quantum corrections to the thermodynamics of charged black holes,'' Phys. Rev. D \textbf{54}(4), 2732 (1996).

\bibitem{MV} M.~Buric and V.~Radovanovic,
``Quantum corrections for the Reisner-Nordstrom black hole,''
Class. Quant. Grav. \textbf{16}, 3937-3951 (1999).
 
\bibitem{Notes} T.~Anegawa and N.~Iizuka,
``Notes on islands in asymptotically flat 2d dilaton black holes,''
JHEP \textbf{07}, 036 (2020). 

\bibitem{HSS} T.~Hartman, E.~Shaghoulian and A.~Strominger,
``Islands in Asymptotically Flat 2D Gravity,''
JHEP \textbf{07}, 022 (2020).


\end{thebibliography}
\end{document}